\documentstyle[12pt,aaspp4]{article}
\newcommand\ltorder{\mathrel{\raise.3ex\hbox{$<$}\mkern-14mu
             \lower0.6ex\hbox{$\sim$}}}
\newcommand\gtorder{\mathrel{\raise.3ex\hbox{$>$}\mkern-14mu
             \lower0.6ex\hbox{$\sim$}}}

\begin{document}

\title{Structure of Dark Matter Halos From Hierarchical Clustering. \\
III. Shallowing of The Inner Cusp}

\author{Toshiyuki Fukushige}

\affil{Department of General Systems Studies,\\
College of Arts and Sciences, University of Tokyo,\\
 3-8-1 Komaba, Meguro-ku, Tokyo 153, Japan}

\author{Atsushi Kawai}

\affil{Faculty of Human and Social Studies,\\
Saitama Institue of Technology,\\
1690 Fusaiji, Okabe, Ohsato, Saitama 369-0293, Japan}
%\affil{Computational Science Division, Advanced Computing Center,\\
%RIKEN (Institute of Physical and Chemical Research),\\
%2-1 Hirosawa, Wako-shi, Saitama, 351-0198, Japan}

\author{Junichiro Makino}

\affil{Department of Astronomy,\\
School of Sciences, University of Tokyo,\\
 7-3-1 Hongo, Bunkyo-ku, Tokyo 117, Japan}

%\centerline{\bf (Version of \today)}

\begin{abstract}

We investigate the structure of the dark matter halo formed in the cold
dark matter scenarios by N-body simulations with parallel treecode on
GRAPE cluster systems.  We simulated 8 halos with the mass of $4.4\times
10^{14}M_{\odot}$ to $1.6\times 10^{15}M_{\odot}$ in the SCDM and LCDM
model using up to 30 million particles.  With the resolution of our
simulations, the density profile is reliable down to 0.2 percent of
the virial radius.  Our results show that the slope of inner cusp within
1 percent virial radius is shallower than $-1.5$, and the radius where
the shallowing starts exhibits run-to-run variation, which means the
innermost profile is not universal. 

\end{abstract}


\section{Introduction}

Since the "finding" of the universal profile by Navarro, Frenk, and
White (1996, 1997, hereafter NFW), the structure of dark matter halos
formed through dissipationless hierarchical clustering from cosmological
initial setting has been explored by many researchers.  NFW performed a
number of $N$-body simulations of the halo formation using 10-20k
particles and found that the profile of dark matter halo could be fitted
to a simple formula (hereafter, NFW profile)
\begin{equation}
\rho={\rho_0 \over (r/r_0)(1+r/r_0)^2}
\label{eqnfw}
\end{equation}
where $\rho_0$ is a characteristic density and $r_0$ is a scale radius. 
They also argued that the profile has the same shape, independent of the
halo mass, the power spectrum of the initial density fluctuation or
other cosmological parameters. 

Several groups reported the results of similar simulations with much
higher resolutions.  However, disagreement concerning the inner
structure still remains.  Some researchers claimed that the slope of the
inner cusp is steeper than that in the NFW's results.  Fukushige and
Makino (1997) performed a similar simulation with 768k particles and
found that the galaxy-sized halo has a cusp steeper than $\rho\propto
r^{-1}$.  Moore et al.  (1998, 1999 hereafter M99) and Ghigna et al. 
(2000) performed simulations with up to 4M particles and obtained the
results that the profile has a cusp proportional to $r^{-1.5}$ both in
galaxy-sized and cluster-sized halos.  M99 proposed the modified
universal profile (hereafter, M99 profile),
\begin{equation}
\rho = {\rho_0 \over (r/r_{\rm 0})^{1.5}[1+(r/r_{\rm 0})^{1.5}]}. 
\label{eqm99}
\end{equation}
Fukushige and Makino (2001, Paper I; 2003, Paper II) performed two
series of $N$-body simulations, and found that
the halos have density cusps proportional to $r^{-1.5}$, independent of
the halo mass and cosmological models. 

On the other hands, other researchers obtained the slope of inner cusp
shallower than $-1.5$ and close to that in the NFW profile.  Jing and
Suto (2000, 2002) performed a series of $N$-body simulations and
concluded that the power of the cusp depends on mass.  It varies from
$-1.5$ for galaxy mass halo to $-1.1$ for cluster mass halo.  Klypin et
al.  (2001) obtained the slope at the center that could be approximated
by $r^{-1.5}$.  They, however, argued that the NFW fit is still good up
to their resolution limit.  Power et al.  (2003) simulated an LCDM
galaxy-sized halo with 3M particles and claimed that their circular
velocity profile obtained is in better agreement with the NFW profile
than with the M99 profile. 

The purpose of this paper is to explore the inner structure of the dark
matter halo by means of $N$-body simulations with about 10 times higher
mass resolution than that of previous simulations.  We simulated the
formation of 8 cluster-sized halos in the SCDM and LCDM models using
parallel Barnes-Hut treecode on parallel GRAPE cluster. 

The structure of this paper is as follows.  In section 2, we describe
the model of our $N$-body simulation.  In section 3, we present the
results of simulation.  Section 4 is for conclusion and discussion. 

%%%%%%%%%%%%%%%%%%%%%%%%%%%%%%%%%%%%%%%%%%%%%%%%%%%

\section{Simulation Method}

We consider two cosmological models, SCDM model ($\Omega_0=1.0$,
$h=0.5$, $\sigma_8=0.6$) and LCDM model ($\Omega_0=0.3$, $\lambda_0=0.7$,
$h=0.7$, $\sigma_8=1.0$).  Here, $\Omega_0$ is the density parameter,
$\lambda_0$ is the dimensionless cosmological constant, and $H_0 = 100 h
\, {\rm km/s}\cdot{\rm Mpc}^{-1}$ at the present epoch.  The amplitudes
of the power spectrum in CDM models are normalized using the top-hat
filtered mass variance at $8 h^{-1}$ Mpc according to the cluster
abundance (Kitayama \& Suto 1997). 
	
We simulate the formation of the dark matter halos using the
"re-simulation" method, which has been a standard method for the
simulation of halo formation since NFW (1996).  The procedure for
setting the initial condition of halos are the same as that used in
Paper II. 

We first performed large scale cosmological simulations with $3.7\times
10^6$ particles in a sphere of 300$h$Mpc comoving radius.  We regard
spherical overdensity regions around local potential minima within
$r_{\rm v}$ as candidate halos.  We define the radius $r_{\rm v}$ such
that the spherical overdensity inside is 178$\Omega_0^{0.3}$ times the
critical density for SCDM and 178$\Omega_0^{0.4}$ times for LCDM model
(Eke, Cole, Frenk 1998). 

We selected 8 halos from the catalog of candidate halos.  The selected
halos are summarized in Table \ref{tab1}.  We selected the three most
massive halos and one halo randomly from halo candidates lying within
200$h$Mpc from the center in both models (so that the external tidal
field can be included).  We express a region within $5r_{v}$ from the
center of the halo at $z=0$ in the cosmological simulation with larger
number of particles.  We place particles whose mass is as same as that
in the cosmological simulation in a sphere of $\sim$ 100$h$Mpc comoving
radius surrounding the high resolution region, in order to express the
external tidal field.  The total number of particles, $N$, is listed in
Table \ref{tab1}.  The generation of initial density fluctuation were
done on HITACHI SR8000 (1 node) in Information Technology Center,
University of Tokyo using COSMICS2 package (Bertschinger 2001). 

\begin{table}[h]
\caption{Run Properties\label{tab1}}
\begin{tabular}{lc|ccc|ccc}
\hline
\hline
Model & Run & $M_{\rm v}$($M_{\odot}$) & $r_{\rm v}$ (Mpc) 
& $N_{\rm v}(\times 10^6)$ & $N(\times 10^6)$ & $m$ ($10^{7}M_{\odot}$)& $1+z_i$ \\
\hline
SCDM & S1 & $1.58\times 10^{15}$ & 3.08 & 29.2 & 60.3 & 5.39 & 44.2 \\
     & S2 & $1.21\times 10^{15}$ & 2.84 & 31.2 & 60.7 & 3.86 & 45.5 \\
     & S3 & $1.21\times 10^{15}$ & 2.84 & 4.5 & 10.0 & 26.5 & 37.9 \\
     & S4 & $4.47\times 10^{14}$ & 2.03 & 6.9 & 13.9 & 6.46 & 43.4 \\
\hline
LCDM & L1 & $9.61\times 10^{14}$ & 2.43 & 25.2 & 62.8 & 3.80 & 51.1 \\
     & L2 & $6.96\times 10^{14}$ & 2.20 & 26.0 & 59.9 & 2.67 & 52.7 \\
     & L3 & $6.49\times 10^{14}$ & 2.15 & 7.2 & 16.7 & 9.01 & 47.5 \\
     & L4 & $4.45\times 10^{14}$ & 1.88 & 7.8 & 13.5 & 5.67 & 49.4 \\
\hline
\end{tabular}
\end{table}

We integrate the system directly in the physical coordinates for both
the cosmological and halo simulations. We 
used a leap-flog integrator with shared and constant timestep.  The step
size for the cosmological simulation is $\Delta t/(t_{\rm H}-t_{\rm
i})=1/1024$ and that for the halo simulation is $1/4096$.  Here, $t_{\rm
H}$ is the Hubble time and $t_{\rm i}$ is the time at which the
simulation starts.  The gravitational softening is constant in the
physical coordinates and the length $\varepsilon_{\rm grav}$ is 5kpc for
the cosmological simulation, and 1kpc for Runs S1, S2, L1, and L2, and
2kpc for other runs of halo simulations. 

The force calculation is done with the parallel Barnes-Hut tree code on
GRAPE clusters (Kawai, Makino 2003)\footnote{The source code for both
serial and parallel implementations are available upon request.}.  GRAPE
is a special-purpose computer designed to accelerate $N$-body
simulations.  The parallelization scheme we used are basically the same
as Warren \& Salmon's (1993) Hashed Oct-Tree algorithm, except that we
incorporated Barnes' (1990) modified algorithm.  The modification is
necessary in order to make GRAPE work efficiently (Makino 1991).  We use
only the dipole expansion and the opening parameter $\theta = 0.4$ for
the cosmological simulation and $\theta=0.5$ for the halo simulation. 

For high-resolution halo simulations, we used both a parallel GRAPE-5
cluster at University of Tokyo and a parallel MDGRAPE-2 cluster at
RIKEN.  The parallel GRAPE-5 cluster consists of 8 host
computers (Pentium 4/1.9GHz, i845) each of which has one GRAPE-5 (Kawai
et al.  2000) board.  The parallel MDGRAPE-2 cluster consists of 8 
host computers (Pentium4/2.2GHz, i850) each of which has one
MDGRAPE-2 (Susukita et al.  2003) board.  For cosmological simulations
we used one board GRAPE-5.  The simulation presented below needs, for
example in Run S2, $\sim 300$ seconds per timestep, and thus one run
(4096 timesteps) is completed in 350 hours (wallclock time) with the
GRAPE-5 cluster. 

%%%%%%%%%%%%%%%%%%%%%%%%%%%%%%%%%%%%%%%%%%%%%%%%%%%

\section{Results}

\subsection{Snapshots}

\begin{figure}
\begin{center}
{\leavevmode
\epsfxsize=17cm
\epsfbox{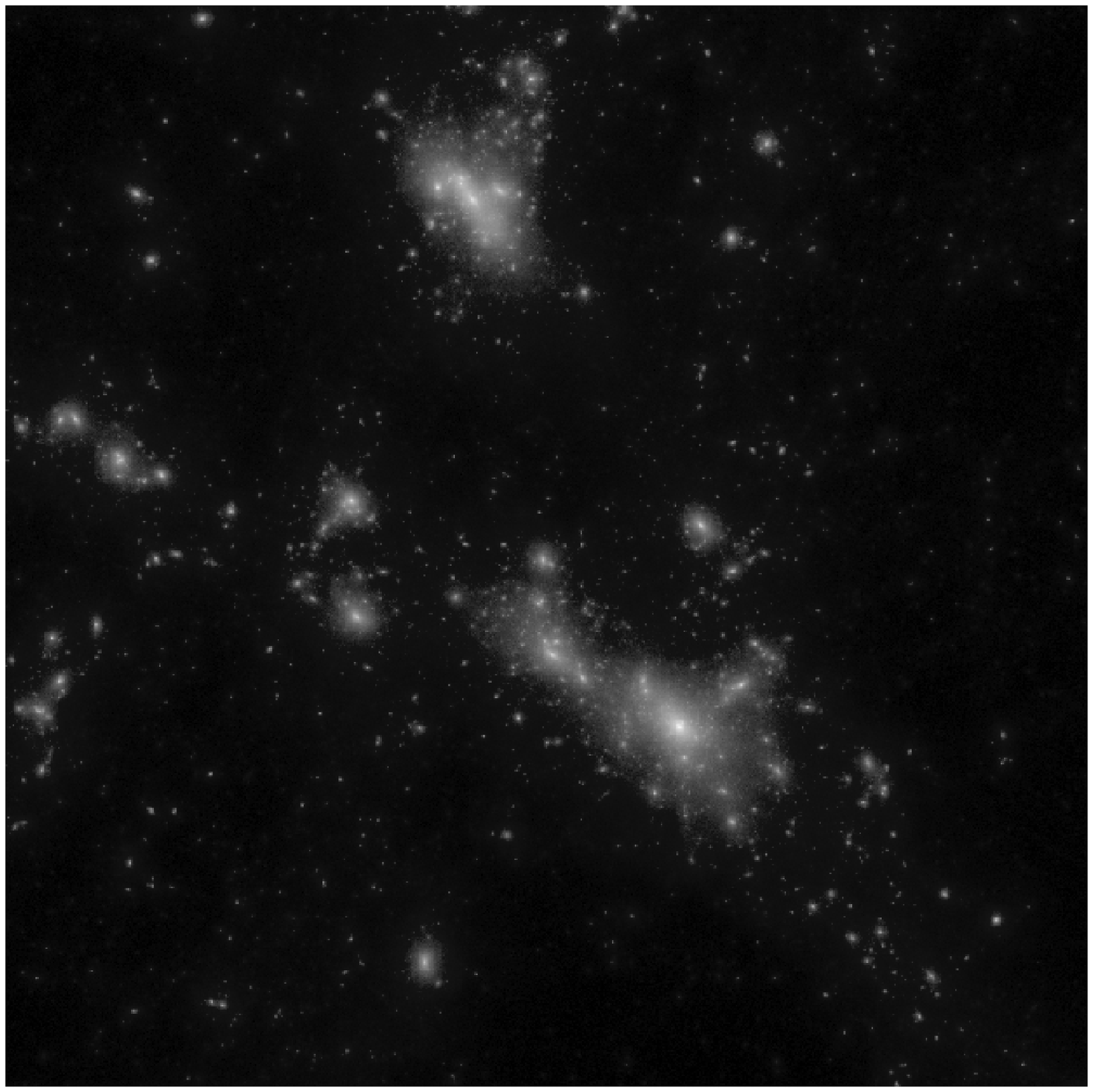}
\caption{Snapplots from Run S2 at $z=0.58$. 
The length of the side is equal to 6.67Mpc. \label{figsnap}}
}
\end{center}
\end{figure}

Figure \ref{figsnap} shows the particle distribution for Run S2 at
$z=0.58$ and $0$.  The length of the side for each panel is 6.67 Mpc. 
For these plots, we shifted the origin of coordinates to the position of
the potential minimum.  In Table \ref{tab1}, we summarized the radius
$r_{\rm v}$, the mass $M_{\rm v}$, and the number of particles $N_{\rm
v}$ within $r_{\rm v}$ at $z=0$. 

\subsection{Density Profile}
\label{secm99}
\label{secac}

Figures \ref{figrhos} and \ref{figrhol} show the density profiles of all
runs at $z=0$ for SCDM and LCDM models, respectively.  The exception is
Run L4, for which we plot the density profile at $z=0.06$ because the
merging process occurs just near the center of halos at $z=0$.  The
position of the center of the halo was determined using the potential
minimum and the density was averaged over each spherical shell whose
width is $\log_{10}(\Delta r)=0.0172$.  For the illustrative purpose,
the densities are shifted vertically. 
 
We plot the densities by the thick (colored magenta in online edition)
lines only if the criteria for two-body relaxation introduced in Paper
I, $t_{\rm rel}(r)/t > 3$, is satisfied, where $t_{\rm rel}(r)$ is the
local two-body relaxation time given by
\begin{equation}
t_{\rm rel}={0.065v^3 \over G^2\rho m \ln(R_{\rm max}/\varepsilon)},
\label{eqtr}
\end{equation}
(cf. Spitzer 1987) 
and $R_{\rm max}$ is a maximum impact parameter.  Here we set $R_{\rm
max}$ to 1 Mpc as a system size.  We also confirmed 
that other numerical artifacts due to the time integration 
did not influence the density profile as will be 
discussed in section \ref{secdt}.  The potential softening does not
influence the profile since the reliability limit obtained by the above
criterion is more than three times larger than the softening length for
all runs. 

At $r > 0.02$ Mpc or $r > 0.01 r_{\rm v}$, the density profiles are in
good agreement to the profile given by equation (\ref{eqm99}) (the M99
profile) in all runs.  This result is consistent with the previous
simulations performed with a few million particles (M99, Ghigna et al. 
2000, Paper I, Paper II).  The fitting here was done using $M_{\rm v}$
and the least square fit of $(\rho-\rho_{\rm M99})/\rho_{\rm M99}$ at
$0.03< r <0.5$ Mpc.  The scale radii $r_0$ obtained by the fitting are
summarized in Table \ref{tab2}. 

\begin{table}[h]
\caption{Fitting Parameters\label{tab2}}
\begin{tabular}{c|cccc|c}
\hline
\hline
Run & $r_0$ (Mpc) & $r_0$ (Mpc) & $r_0$ (Mpc) & $r_{\rm c}$ (Mpc) & $r_{\rm v}$ (Mpc) \\
    &  for $\rho_{\rm M99}$  & for $\rho_{\rm NFW}$ & for $\rho_{\rm n1}$
& for $\rho_{\rm n2}$ \\
\hline
S1 & 1.36 & 0.41 & 0.70 & 0.0014 & 3.05\\
S2 & 1.31 & 0.40 & 0.68 & 0.0014 & 2.82\\
S3 & 0.88 & 0.42 & 0.60 & 0.0036 & 2.82\\
S4 & 0.66 & 0.29 & 0.44 & 0.0023 & 1.97\\
L1 & 0.75 & 0.31 & 0.50 & 0.0023 & 2.40\\
L2 & 0.95 & 0.33 & 0.52 & 0.0015 & 2.13\\
L3 & 0.48 & 0.23 & 0.34 & 0.0027 & 2.13\\
L4 & 0.57 & 0.26 & 0.38 & 0.0024 & 1.82\\
\hline
\end{tabular}
\end{table}

On the other hands, at $ r < 0.01 r_{\rm v}$, we can see the shallowing
of the cusp from the power $-1.5$ for all run.  The degree of the
shallowing seems to increase as the radius decreases, which seemingly
suggests that the inner cusp profile does not converge to a single
slope.  Moreover, the point where the profile starts to depart from the
$r^{-1.5}$ cusp is different for different runs.  For example, in Run S1
the departure starts at $\sim 0.005 r_{\rm v}$, while at $\sim 0.02
r_{\rm v}$ in Run L3.  This means that {\it the density profile is not
universal}. 

In Figure \ref{figuni} we plot the density profiles for all runs scaled
by $r_0$ and $\rho_0$, together with the M99 profile.  We can see that
at $r/r_0 < 0.05$ all profiles are systematically shallower than the
$r^{-1.5}$ cusp, and that in this region run-to-run variation of the
profile is significant.  On the other hands, at $r/r_0 > 0.05$ the
profiles are in good agreement with M99 profile.  Although there are
some dispersion from M99 profile at $r/r_0>0.3$, they are not
systematic. 

\subsection{Reliability}

\subsubsection{Two-body relaxation}

We test the reliability of the criterion (\ref{eqtr})using the
simulations of the same initial condition as used in Run S1 but with
several different values for the total number of particles ($N$). 
Except for $N$, we used the same simulation parameters as in Run S1. 
Figure \ref{fignevtb} show the cumulative mass $M_{\rm r}(r)$ within the
radii, 0.1, 0.03, 0.01, 0.005, and 0.003 Mpc, as a function of time, for
three simulations with 29(Run S1), 14 and 1 million particles within
$r_{\rm v}$.  Figure \ref{figrhoevtb} shows the final density profiles for
three simulations.  The vertical bars indicates the reliability limit
obtained by the criterion (\ref{eqtr}). 

In Figure \ref{fignevtb}, we can see that the cumulative mass evolution
obtained in the simulations with 29 and 14 millions particle are in good
agreement for $r>0.01$ Mpc.  This agreement indicates that our criterion
(0.009 Mpc for 14 millions particle run) gives a good reliability limit. 
In Figure \ref{figrhoevtb} we can also see that the density $\rho$
obtained in the simulations with 29 and 14 millions particle are in good
agreement outside the reliability limit of 14 millions particle run
(0.009 Mpc).  The agreement of the averaged density is somewhat worse
than that of the density.  This is because the averaged density is
integrated quantity.  Any error of the density inside the sphere of
radius $r$ affects the average density at radius $r$. 

Recently, Power et al.  (2003) proposed another reliability criterion
for the two-body relaxation, given by
\begin{equation}
{t_{\rm rel}(r)\over t} = \displaystyle{{N(r)\over 8\ln{N(r)}}
\left({\rho_{\rm ave}\over 200\rho_{\rm crit}}\right)^{-1/2}} > 0.6. 
\end{equation}
Although their function form is different from ours and ignores the
dependence on potential softening (see, Fig 3 of Paper I), it gives
reliability limits similar to ours.  For example, in Run S1, the
reliability limit given by their criterion is 0.007, 0.009 and 0.025 Mpc
for simulations with 29, 14 and 1 millions particles.  These values are
within 15\% of our limit shown in Figure \ref{figrhoevtb}. 

\subsubsection{Time integration}
\label{secdt}

If the stepsize for the time integration is too large, it also influences
the profile.  We check whether the stepsize of time integration used in
our simulations is small enough by performing the simulations from the
same initial model as Run S1 but with several different stepsize
($\Delta t$).  Except for $\Delta t$, we used the same simulation
parameters as in Run S1.  Figure \ref{fignevdt} show the cumulative mass
within the radii, 0.1, 0.03, 0.01, 0.005, and 0.003 Mpc, as a function
of time for three simulations with $\Delta t/(t_{\rm H}-t_{\rm
i})=$1/4096(Run S1), 1/2048 and 1/1024.  Figure \ref{figrhoevdt} show
the profile of the density $\rho$ for three simulations at $t/(t_{\rm
H}-t_{\rm i})=0.78125$.  We plot the profile at this time since, in the
simulation with $\Delta t/(t_{\rm H}-t_{\rm i})=$1/2048, merging process
occurs near the center of halos at around $z=0$. 

In these figures we can see that larger stepsize makes the central
profile shallower.  The density profile outside of 0.007 Mpc converges
even by adapting 1/2048.  Therefore, we can conclude that the stepsize
of $\Delta t/(t_{\rm H}-t_{\rm i})=$1/4096 did not introduce any
numerical artifact. 

Power et al.  (2003) investigated influences of the large stepsize on
the profile, and showed that the influence depends also on the softening
length.  They found that if potential softening length is larger than an
optimal length, $\varepsilon \simeq 4r_{\rm v}/\sqrt{N_{\rm v}}$, a
reliability limit is given by
\begin{equation}
{t_{\rm c}(r)\over t_{\rm c}(r_{\rm v})} = 15\displaystyle{\left({\Delta
t\over t_0}\right)^{5/6}}, 
\label{eqdt} 
\end{equation} 
and if softening length is smaller than the optical length more
timesteps are required than that given by criterion (\ref{eqdt}). 

However, an application of Power et al.  (2003)'s criterion to our
simulation results seems to give unphysically reliability limits.  For
example, in Run S1, the reliability limit given by criterion
(\ref{eqdt}) is 0.016, 0.038 and 0.083 Mpc for simulations with 1/4096,
1/2048 and 1/1024, respectively.  From Figure \ref{figrhoevdt}, it is
clear that these values are far too large.  Although we do not fully
understand the origin of the difference, such difference is possible. 
The error in time integration is very complicated because it depends not
only on the stepsize and softening length, as Power et al.  (2003)
showed, but also on the integration scheme selected (ex.  variable
timestep or not, in comoving space or in physical space).

\subsection{Fitting by NFW profile}
\label{secnfw}

In Figure \ref{figfitn}, we fit the density profiles for all runs to the
NFW profile.  The fitting here was done using $M_{\rm v}$ and the least
square fit of $(\rho-\rho_{\rm NFW})/\rho_{\rm NFW}$ at $r <0.5$ Mpc
(down to the reliability limit).  The scale radii $r_0$ obtained by the
fitting are summarized in Table \ref{tab2}.  We can see that the NFW
profile is not in good agreement with simulation results, except for Run
L3.  Figure \ref{figres} show the residual, $(\rho-\rho_{\rm
NFW})/\rho_{\rm NFW}$, together with that for the M99 profile.  The
agreement with the NFW profile is not good in all radii, while that with
the M99 profile is not good only in inner region ($r<0.03$Mpc). 
Moreover, we can see that the sign of the residuals for NFW profile
systematically change, which means the central slope of the NFW profile
is too shallower. 

\subsection{Evolution}
\label{secev}

Figures \ref{figevs} and \ref{figevl} show the growth of the density
profile for all runs.  The virial radii and the masses within the virial
radius at the redshift plotted are summarized in Table \ref{tab3}.  We
fit these profiles to the M99 profile.  The fitting procedure is as same
as that for Figure \ref{figrhos}.  The scale radii $r_0$ obtained by the
fitting are summarized in Table \ref{tab3}. 

\begin{table}[h]
\caption{Parameters at higher redshift\label{tab3}}
\begin{tabular}{cc|ccc}
\hline
\hline
Run & $z$ & $r_{\rm v}$ (Mpc) & $M_{\rm v}$ ($M_{\odot}$) & $r_{0}$(Mpc)\\
\hline
S1 & 3.3 & 0.16 & $1.9\times 10^{13}$ & 0.085 \\
   & 1.2 & 0.65 & $1.6\times 10^{14}$ & 0.25 \\
\hline
S2 & 3.3 & 0.15 & $1.6\times 10^{13}$ & 0.056 \\
   & 1.2 & 0.55 & $9.7\times 10^{13}$ & 0.20 \\
\hline
S3 & 1.2 & 0.55 & $9.8\times 10^{13}$ & 0.29 \\
\hline
S4 & 1.2 & 0.46 & $5.5\times 10^{13}$ & 0.20 \\
\hline
L1 & 4.5 & 0.11 & $1.4\times 10^{13}$ & 0.11 \\
   & 1.9 & 0.42 & $1.2\times 10^{14}$ & 0.20 \\
\hline
L2 & 4.5 & 0.086 & $7.3\times 10^{12}$ & 0.05 \\
   & 1.9 & 0.40 & $7.0\times 10^{13}$ & 0.18 \\
\hline
L3 & 1.9 & 0.32 & $5.4\times 10^{13}$ & 0.24 \\
\hline
L4 & 1.9 & 0.31 & $4.7\times 10^{13}$ & 0.13 \\
\hline
\end{tabular}
\end{table}

At the inner region ($r<0.03$ Mpc), we can see the density keeps almost
unchanged from relatively higher redshift for all runs.  This fact also
can be seen in the evolution of the cumulative mass shown in Figure
\ref{fignevtb}.  This means that the density at the inner region is
determined by that of the smaller halo that collapsed at higher
redshift. 

The density profile of the outer region is formed as the halo grows and
shows universality.  Moreover, the agreement with the M99 profile at
higher redshift is very well down to the radius at which the cusp
shallowing can be seen at $z=0$, independent of the cosmological model
we simulated in this paper.  Figure \ref{figr0} shows the relation
between the scale radius $r_{0}$ and density $\rho_{0}$ obtained by the
fitting.  We can see clearly an evolutionary pass along a line, $\rho_0
\propto r_0^{-1.5}$, also independent of the cosmological model. 

\subsection{Different Fitting}

In section \ref{secm99}, we see that the agreement with the M99 profile
is not good at the inner region ($r<0.02$Mpc), and also that with the
NFW profile is worse in the whole range of profile in section
\ref{secnfw}.  Therefore, it is worthwhile to fit the results to other
profiles.  Here, we try to fit the results to two different profiles. 

Firstly, we fit the results to a profile that has an inner cusp
shallower than that of the M99 profile and steeper than the NFW profile
[fitting (1)], given as
\begin{equation} 
\rho_{\rm n1} = \displaystyle{\rho_{0} \over 
(r / r_{\rm 0})^{\alpha} \left[1+(r / r_{\rm 0})^{3-\alpha}\right]}
\label{eqprof1}
\end{equation} 
where $\alpha$ is the power of the inner cusp and we set $\alpha=1.3$. 
In Figure \ref{figfits}, we fit the density profiles to the profile
given by equation (\ref{eqprof1}).  The fitting here was done using
$M_{\rm v}$ and the least square fit of $(\rho-\rho_{\rm n1})/\rho_{\rm
n1}$ for $r <0.5$ Mpc (down to the reliability limit).  
The scale radii $r_{\rm 0}$ obtained by the
fitting are summarized in Table \ref{tab2}.  Figure
\ref{figresnew} shows the residual $(\rho-\rho_{\rm n1})/\rho_{\rm n1}$. 
The agreement is better than both the M99 and NFW profiles. 

We also tried to add another power law region ($\propto r^{\beta}$) 
to the M99 profile [fitting (2)], given as 
\begin{equation} 
\rho_{\rm n2} = \displaystyle{\rho_{0} \over 
C_0(r / r_{\rm c})^{\beta}
\left[1+(r / r_{\rm c})\right]^{1.5-\beta}
\left[1+(r / r_{\rm 0})^{1.5}\right]}
\label{eqprof2}
\end{equation} 
where
\begin{equation} 
\displaystyle{1 / C_0} = 
(r_0 / r_{\rm c})^{\beta}
\left[1+r_0 / r_{\rm c}\right]^{1.5-\beta},
\end{equation}
$r_{\rm c}$ is another scale radius.  Although this profile includes
more parameters to fit, it is based on the observation that two
different mechanisms might be working in the growth of the halo as
suggested by the analyses in section \ref{secev}. 

In Figure \ref{figfits}, we fit the density profiles to the profile
given by equation (\ref{eqprof2}).  Here, for simplicity, we set
$\beta=0$ for all runs and, therefore, the equation
(\ref{eqprof2}) becomes
\begin{equation} 
\rho_{\rm n2} = \displaystyle{\rho_{0} \over 
C_0 \left[1+(r / r_{\rm c})\right]^{1.5}
\left[1+(r / r_{\rm 0})^{1.5}\right]}
\label{eqnew}
\end{equation} 
where
\begin{equation} 
\displaystyle{1 / C_0} = \left[1+(r_0 / r_{\rm c})\right]^{1.5}
\end{equation}
The fitting here was done using $r_0$ obtained in the fitting to the M99
profile and the least square fit of $(\rho-\rho_{\rm n2})/\rho_{\rm
n2}$ at $r <0.5$ Mpc.  The scale radii $r_{\rm c}$ obtained 
are summarized in Table \ref{tab2}.  Figure
\ref{figresnew} shows the residual, $(\rho-\rho_{\rm n2})/\rho_{\rm n2}$.  As
a matter of course, agreement is better than that for any other
profile, since we increased the number of fitting parameters. 

Unfortunately, in the present simulations, the region that we can use to
determine which fitting formula is more appropriate is not so large. 
Further studies with simulations with higher resolution and larger
number of samples would be necessary. 

\section{Conclusion and Discussion}

We performed $N$-body simulations of dark matter halo formation in SCDM
and LCDM models.  We simulated 8 halos whose mass range is $4.4\times
10^{14}M_{\odot}$ to $1.6\times 10^{15}M_{\odot}$ using up to 30 millions
particles. 

Our main conclusions are: 

\begin{itemize}

\item[(1)] 
We found that, in all runs, the slope of inner cusp within $0.01r_{\rm
v}$ is shallower than $-1.5$, and the radius where the shallowing starts 
exhibits run-to-run variation, which means the profile is not universal. 

\item[(2)] 
We found that the profile is in agreement with the M99 profile for 
$r>0.01r_{\rm v}$, and are not in agreement with the NFW
profile.  We present different fitting formulae to describe the whole
range of the simulation results. 

\end{itemize}

Although we found interesting features in the inner structure of dark
matter halo by new simulations with much higher resolution, we could not
achieve the final understanding of the structure.  One question remained
is whether the CDM halo has a flat core or not.  Another question is
whether the same shallowing can be seen in the halo of galaxy or dwarf
galaxy size.  The origin of the inner structure is also still unclear. 
In order to answer these questions, we are now planning to perform
larger simulations using new GRAPE cluster system. 

\acknowledgements

We are grateful to Yasushi Suto and Atsushi Taruya for many helpful
discussions.  We gratefully acknowledge the use of the initial condition
generator in the publicly available code {\it COSMICS2} developed by E. 
Bertschinger.  We would like to thank all people who have made
contribution to the development of the MDGRAPE-2 system at RIKEN.  A
part of numerical computations were carried out on the GRAPE system at
ADAC (the Astronomical Data Analysis Center) of the National
Astronomical Observatory, Japan.  This research was partially supported
by the Research for the Future Program (JSPS-RFTP 97P01102) and by the
Grants-in-Aid (14740127 and 13440058) of Japan Society for the Promotion
of Science .  Part of this work is carried out while A.K.~is a special
postdoctal researcher of RIKEN.

\newpage

\input epsf.sty

\begin{figure}
\begin{center}
{\leavevmode
\epsfxsize=17cm
\epsfbox{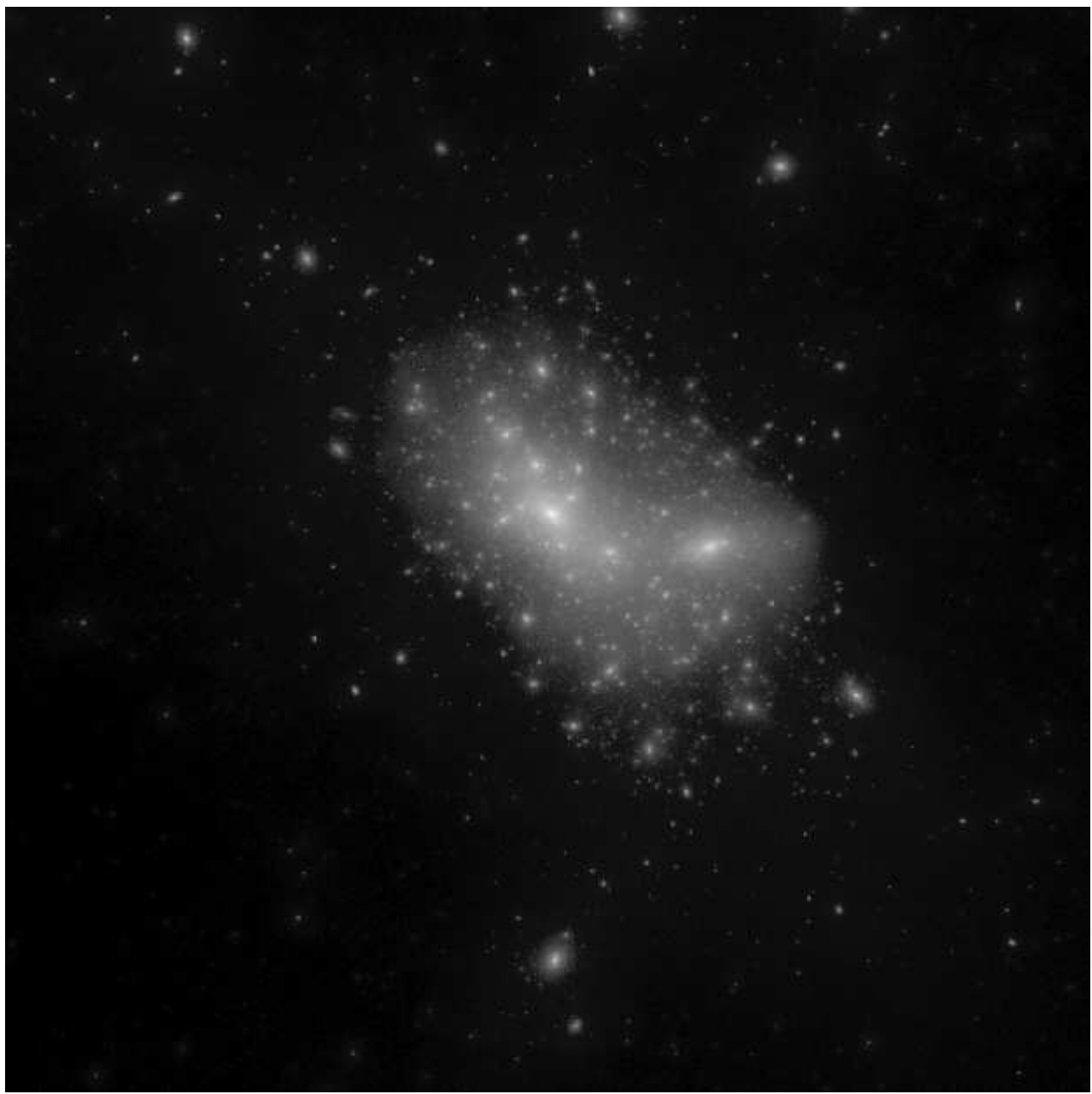}
\caption{Same as Figure \ref{figsnap}, but at $z=0$. \label{figsnap2}}
}
\end{center}
\end{figure}

\begin{figure}
\begin{center}
{\leavevmode
\epsfxsize=17cm
\epsfbox{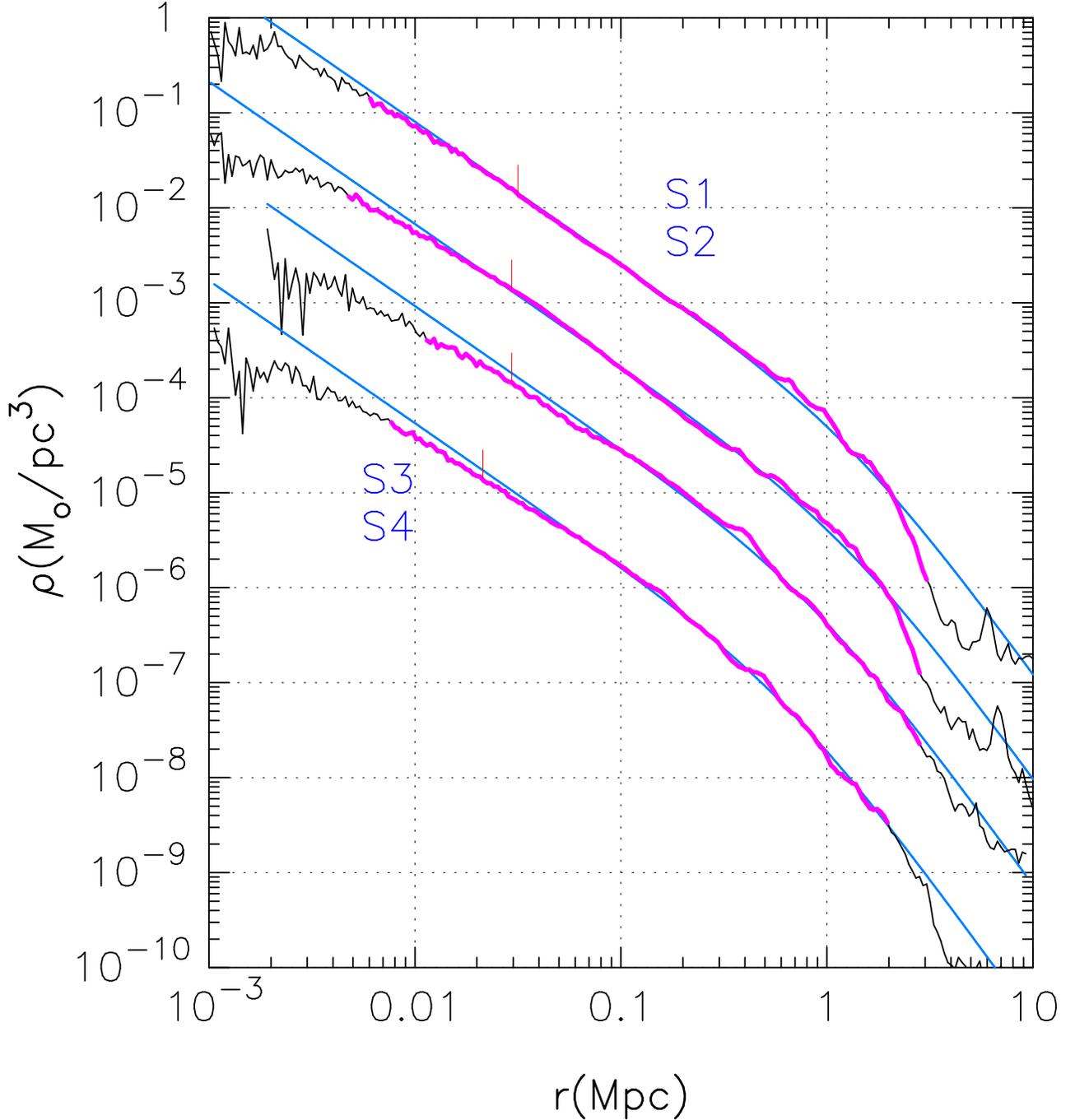}
\caption{
Density profile of the halos for all runs of the SCDM model at $z=0$. 
Only the densities plotted in the thick lines (colored magenta in online
edition) satisfy criterion (\ref{eqtr}) in section \ref{secac} at $r<r_{\rm v}$. 
The labels indicate the run name.  The profiles except for Run S1 are
vertically shifted downward by 1, 2, 3 dex for Runs S2, S3, and S4,
respectively.  The vertical bar above the profiles indicate $0.01r_{\rm
v}$.  The solid curves (colored blue) indicate the density profile given by
equation (\ref{eqm99}) (M99 profile). 
\label{figrhos}}
}
\end{center}
\end{figure}

\begin{figure}
\begin{center}
{\leavevmode
\epsfxsize=17cm
\epsfbox{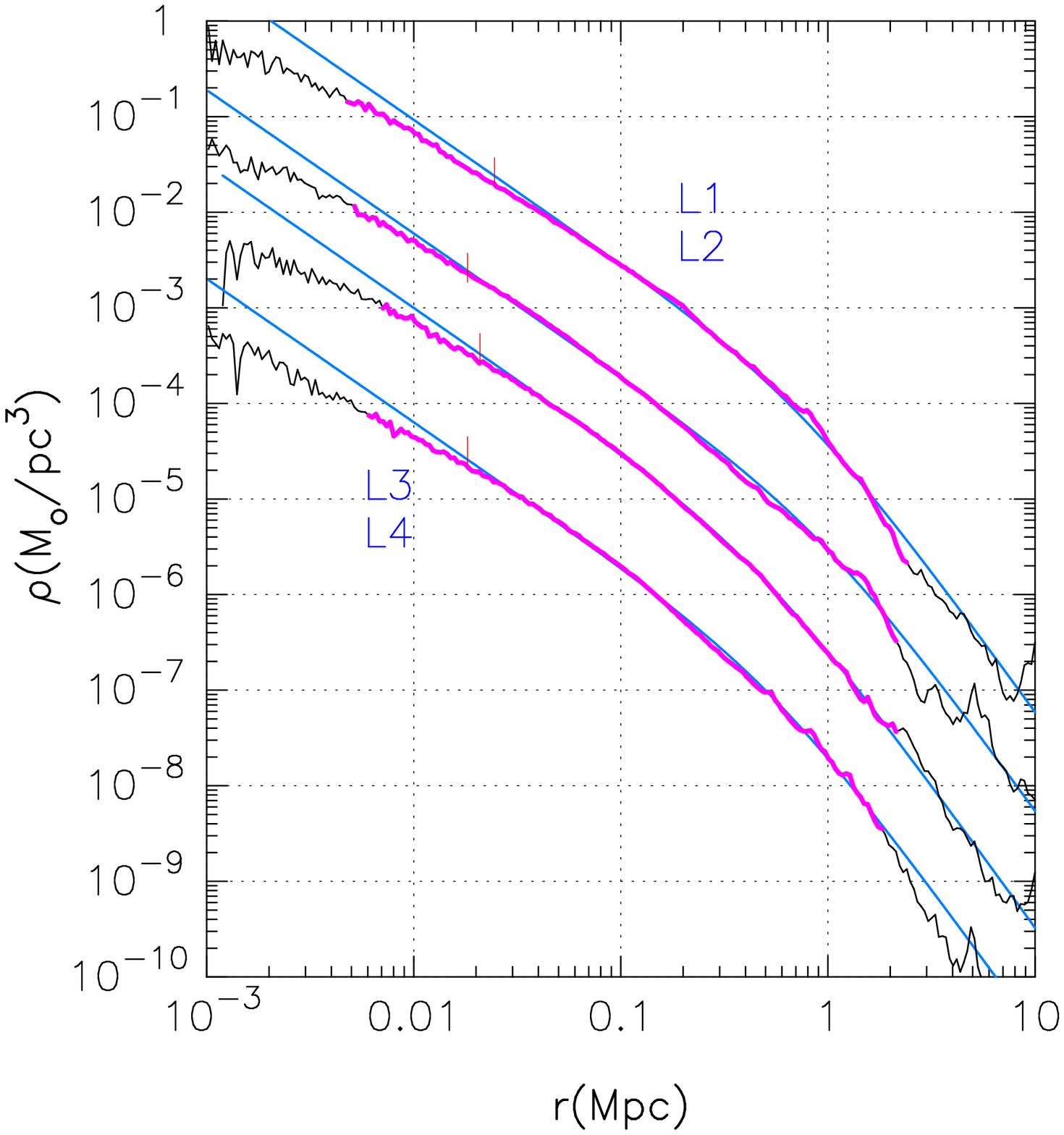}
\caption{
Same as Figure \ref{figrhos}, but for the LCDM model. 
\label{figrhol}}
}
\end{center}
\end{figure}

\begin{figure}
\begin{center}
{\leavevmode
\epsfxsize=17cm
\epsfbox{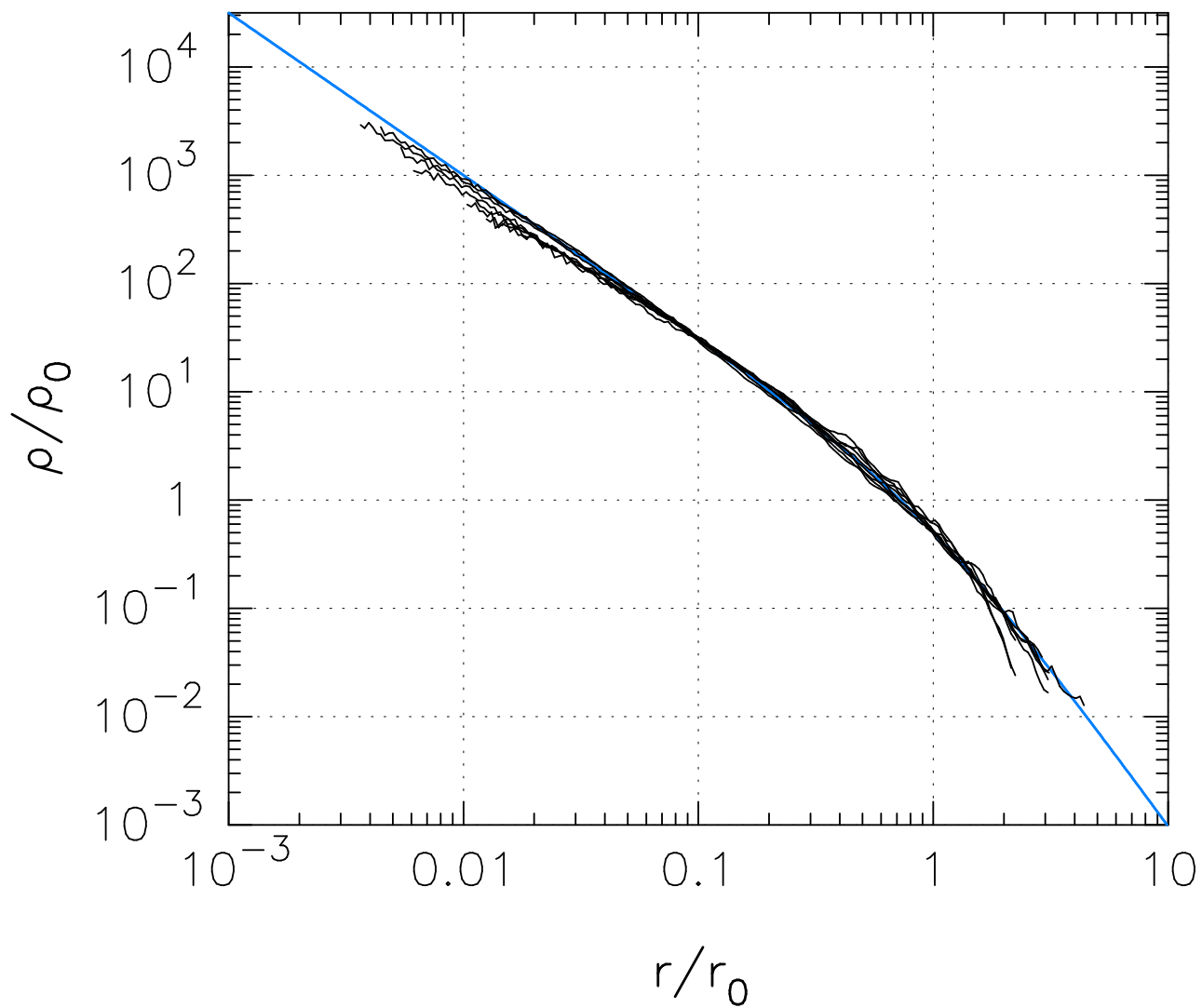}
\caption{
Density profiles for all runs scaled by $r_0$ and $\rho_0$ (Table
\ref{tab2}).  The solid (colored blue in online edition) curves
indicates the density profile given by equation (\ref{eqm99}) (M99
profile). 
\label{figuni}}
}
\end{center}
\end{figure}

\begin{figure}
\begin{center}
{\leavevmode
\epsfxsize=15cm
\epsfbox{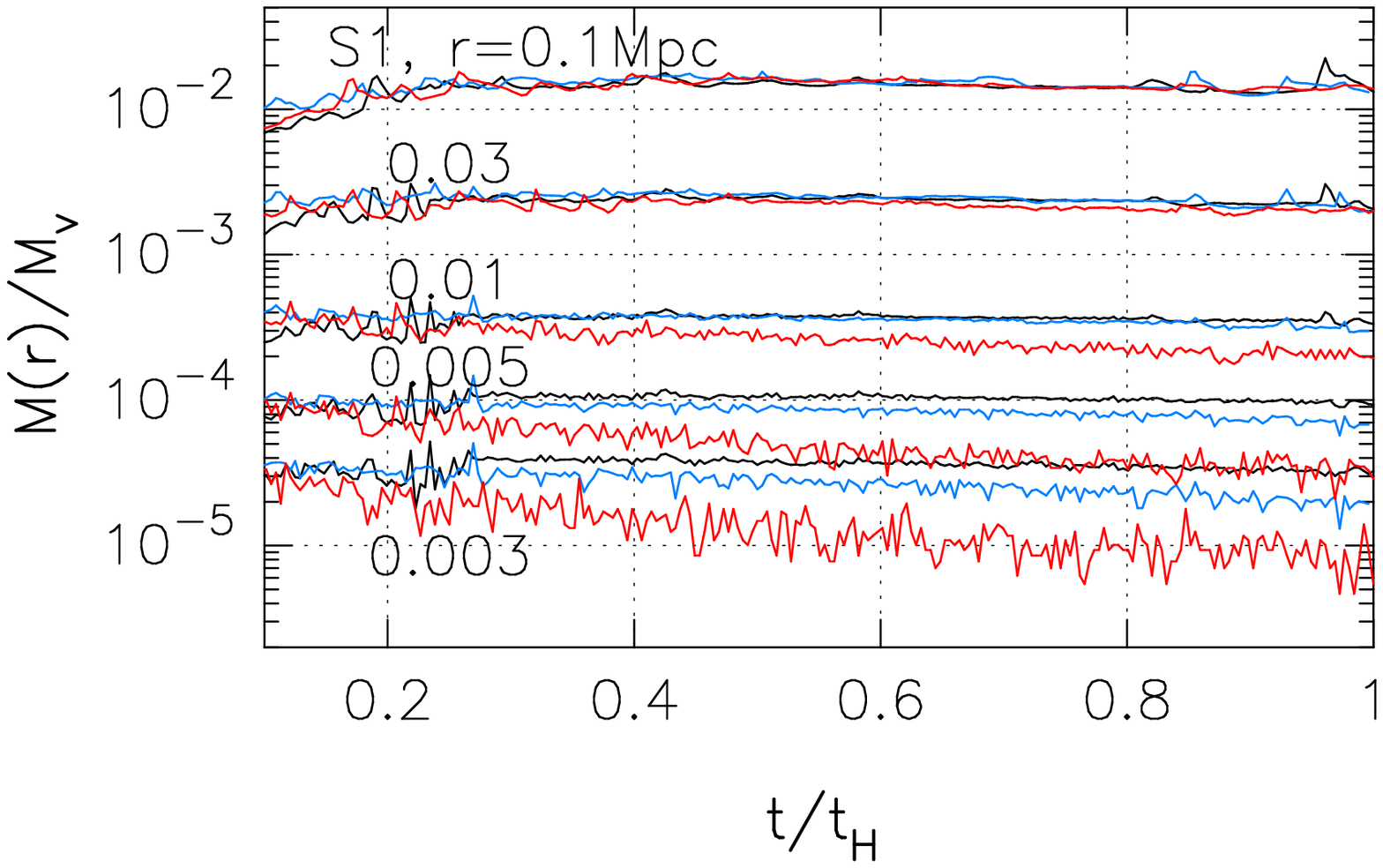}
\caption{Cumulative mass within radii, 0.1, 0.03, 0.01, 0.005, and 0.003
Mpc, as a function of time, for three simulations with 29(Run S1, thick
lines), 14(intermediate thick, colored
blue in online edition) and 1 (thin, colored red) million particles. 
\label{fignevtb}}
}

{\leavevmode
\epsfxsize=12cm
\epsfbox{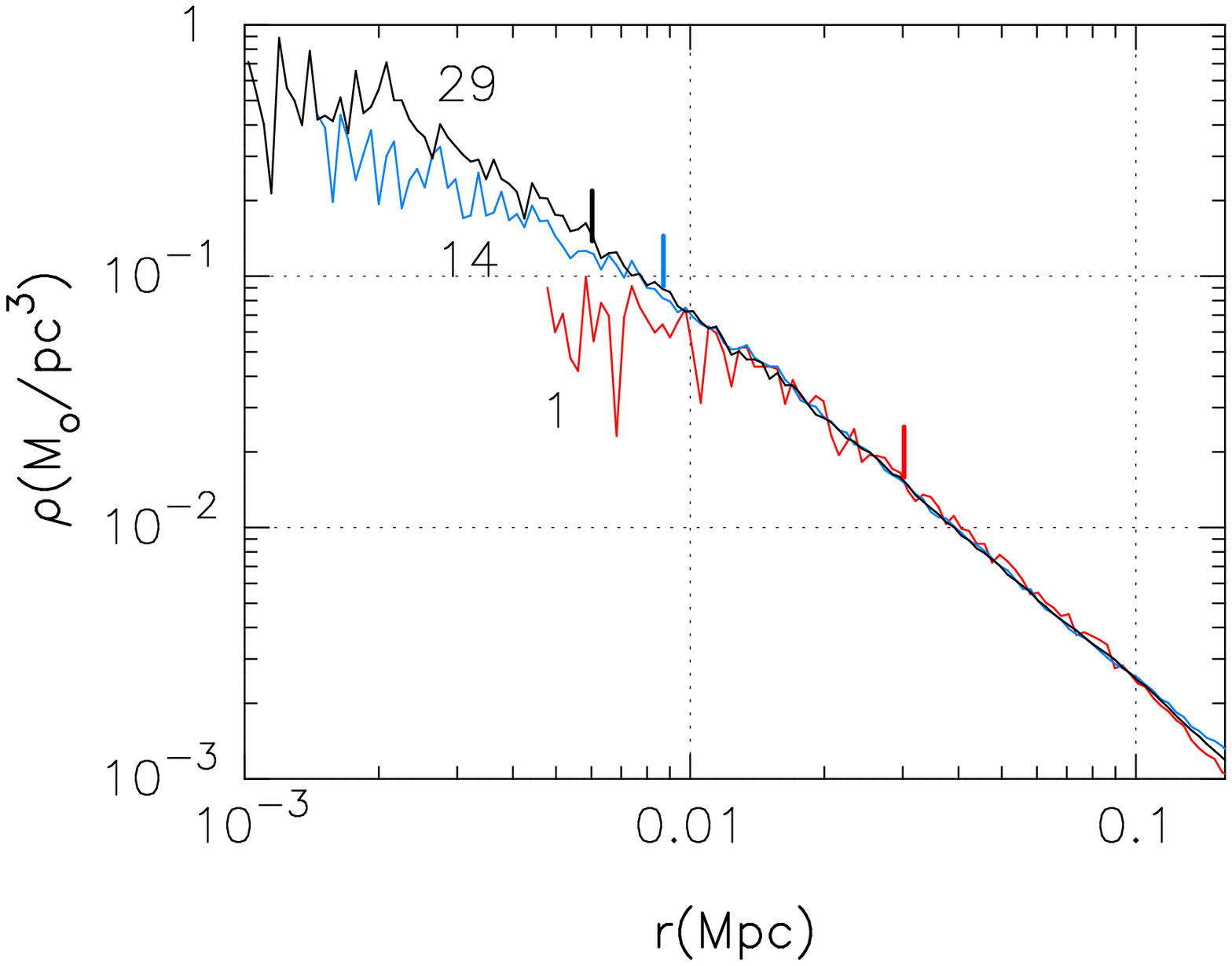}
\caption{Density profiles for three simulations with 29(Run S1, thick lines), 
14(intermediate thick, colored blue in online edition) and 1 (thin, colored red)
million particles.  The vertical bars indicate the reliability limits
obtained by criterion (\ref{eqtr}) in section \ref{secac}. 
\label{figrhoevtb}}
}
\end{center}
\end{figure}

\begin{figure}
\begin{center}
{\leavevmode
\epsfxsize=15cm
\epsfbox{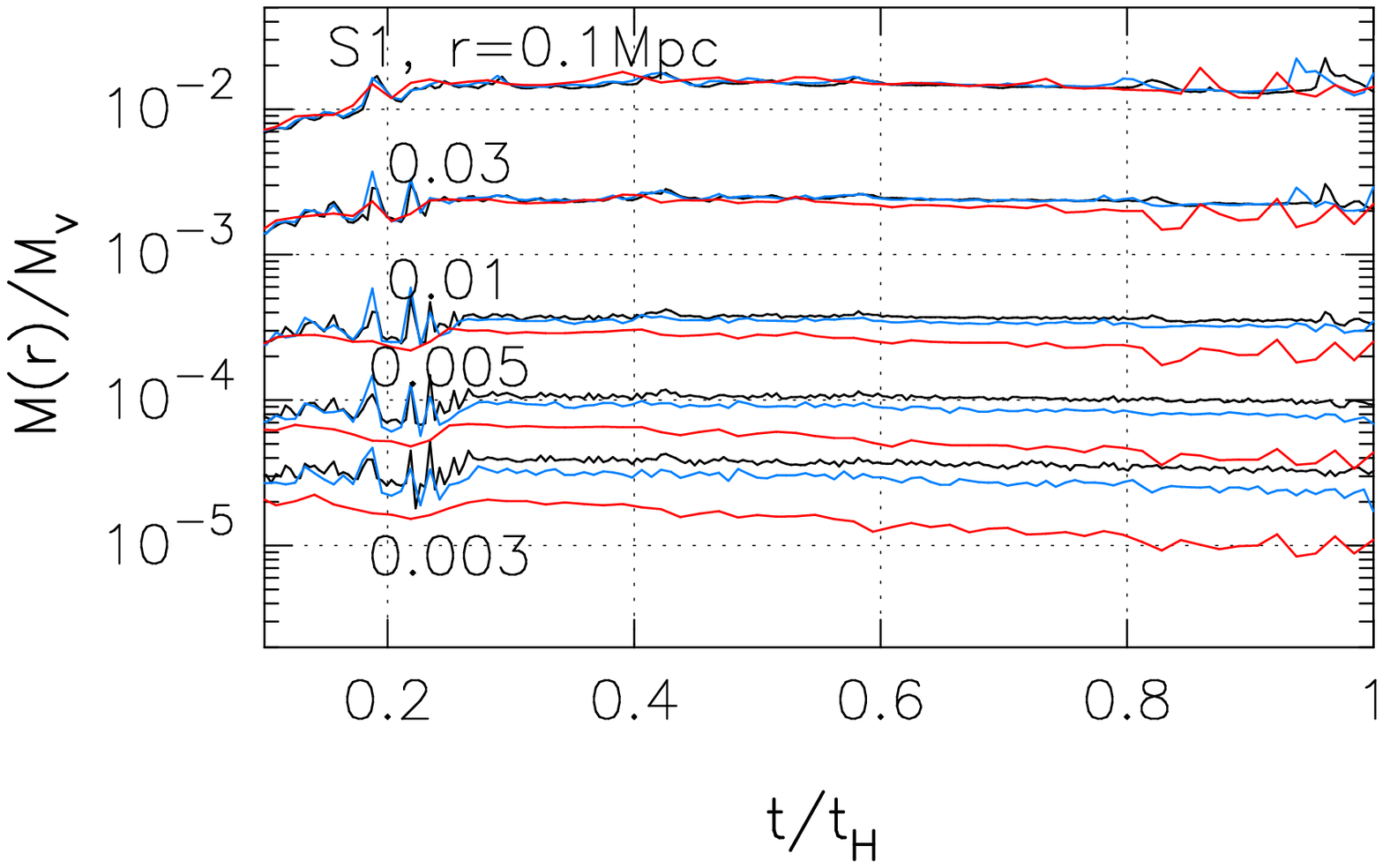}
\caption{
Same as Figure \ref{fignevtb}, but for for three simulations with
$\Delta t/(t_{\rm H}-t_{\rm i})=$1/4096(Run S1, thick lines)
, 1/2048(intermediate thick, colored blue in online edition) and
1/1024(thin, colored red). 
\label{fignevdt}}
}
{\leavevmode
\epsfxsize=12cm
\epsfbox{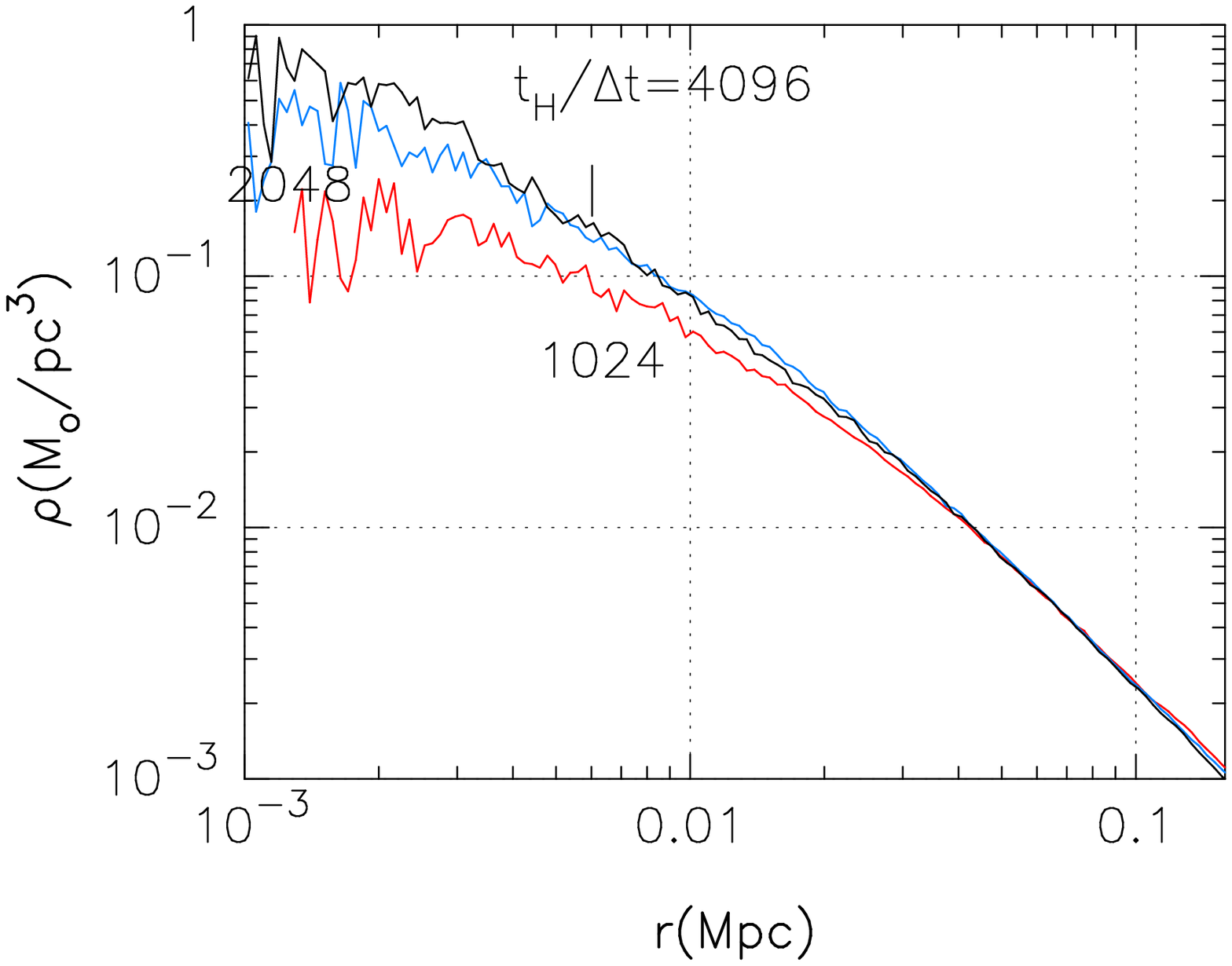}
\caption{
Density profiles for three simulations with $\Delta t/(t_{\rm H}-t_{\rm
i})=$1/4096 (Run S1, thick lines), 1/2048(intermediate thick, colored
blue in online edition) and 1/1024(colored red). 
The vertical bars indicate the reliability limits
obtained by criterion (\ref{eqtr}) in section \ref{secac}. 
\label{figrhoevdt}}
}
\end{center}
\end{figure}

\begin{figure}
\begin{center}
{\leavevmode
\epsfxsize=15cm
\epsfbox{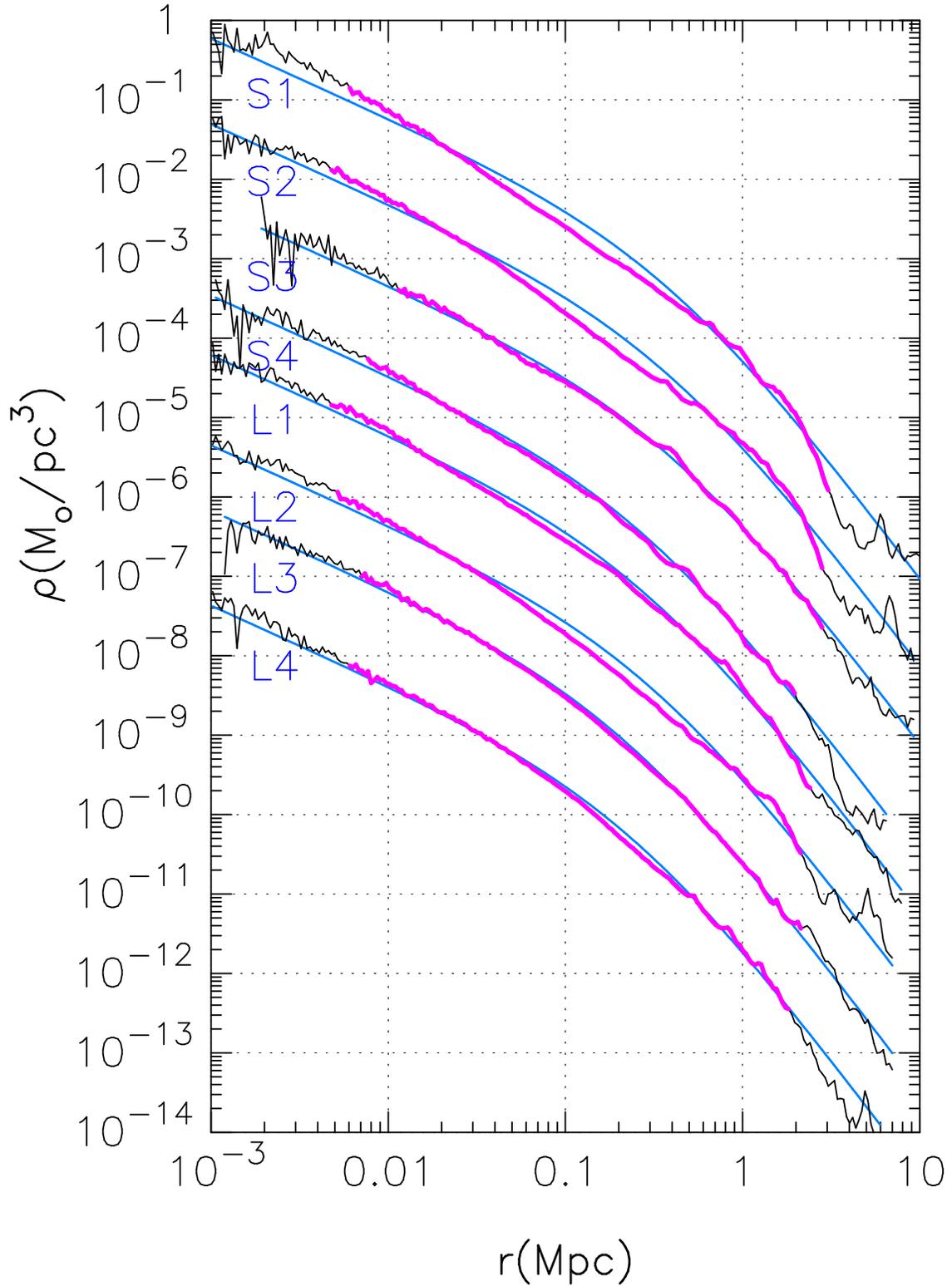}
\caption{
Density profiles for all runs. 
The solid curves (colored blue in online edition) 
indicate the density profile given by equation 
(\ref{eqnfw}) (NFW profile). 
\label{figfitn}}
}
\end{center}
\end{figure}

\begin{figure}
\begin{center}
{\leavevmode
\epsfxsize=15cm
\epsfbox{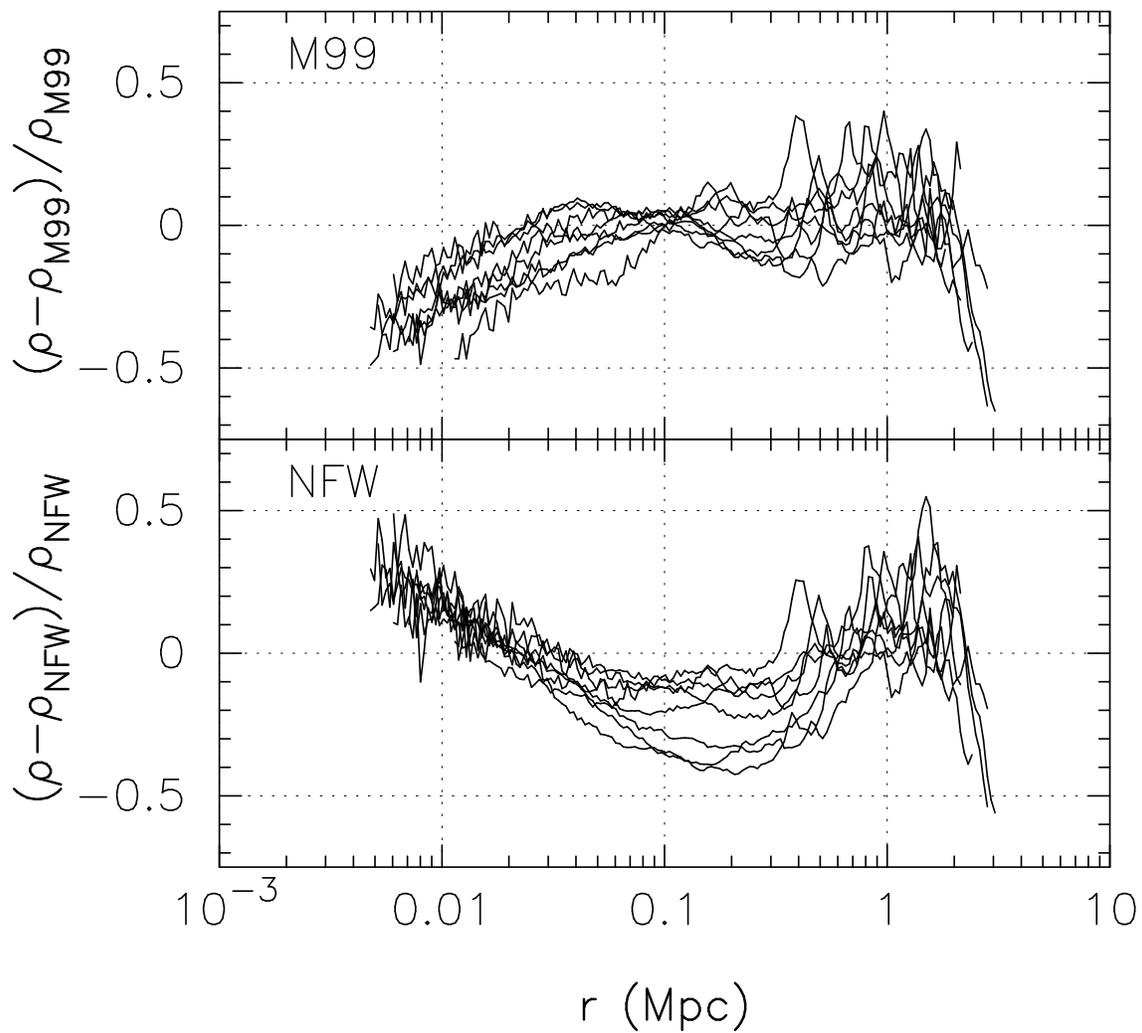}
\caption{
Residuals $(\rho-\rho_{\rm M99})/\rho_{\rm M99}$ 
and $(\rho-\rho_{\rm NFW})/\rho_{\rm NFW}$ 
as a function of radius. 
\label{figres}}
}
\end{center}
\end{figure}

\begin{figure}
\begin{center}
{\leavevmode
\epsfxsize=15cm
\epsfbox{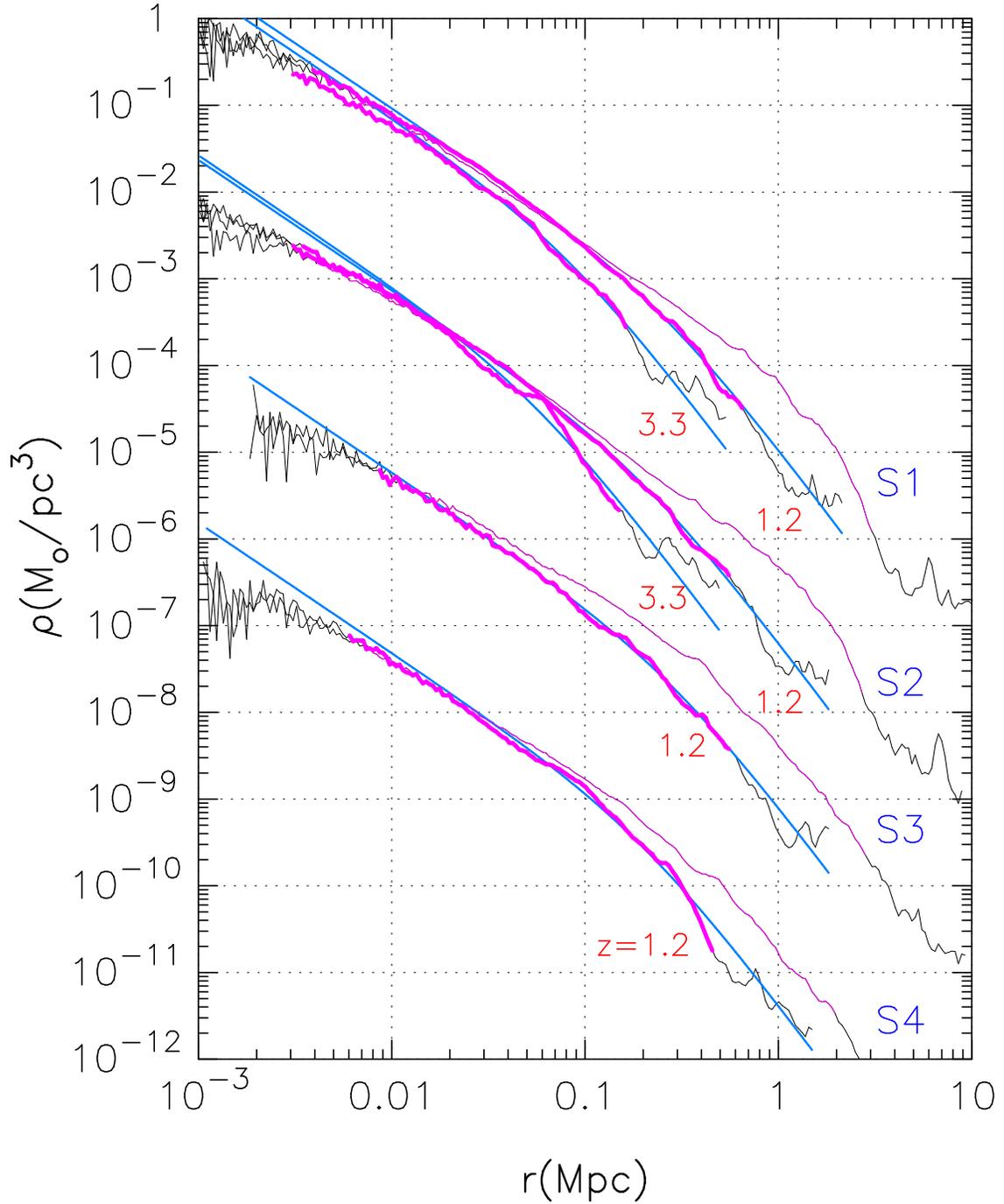}
\caption{
Evolution of density profile for all runs of the SCDM model.  The
numbers near profiles indicate the redshift.  The profiles at $z=0$ are
plotted by the thin lines.  Only the densities plotted in the thick 
(colored magenta in online edition) lines satisfy criterion (\ref{eqtr}) in
section \ref{secac} at $r<r_{\rm v}$.  The solid curves (colored blue)
indicate the density profile given by equation (\ref{eqm99}) (M99
profile). 
\label{figevs}}
}
\end{center}
\end{figure}

\begin{figure}
\begin{center}
{\leavevmode
\epsfxsize=15cm
\epsfbox{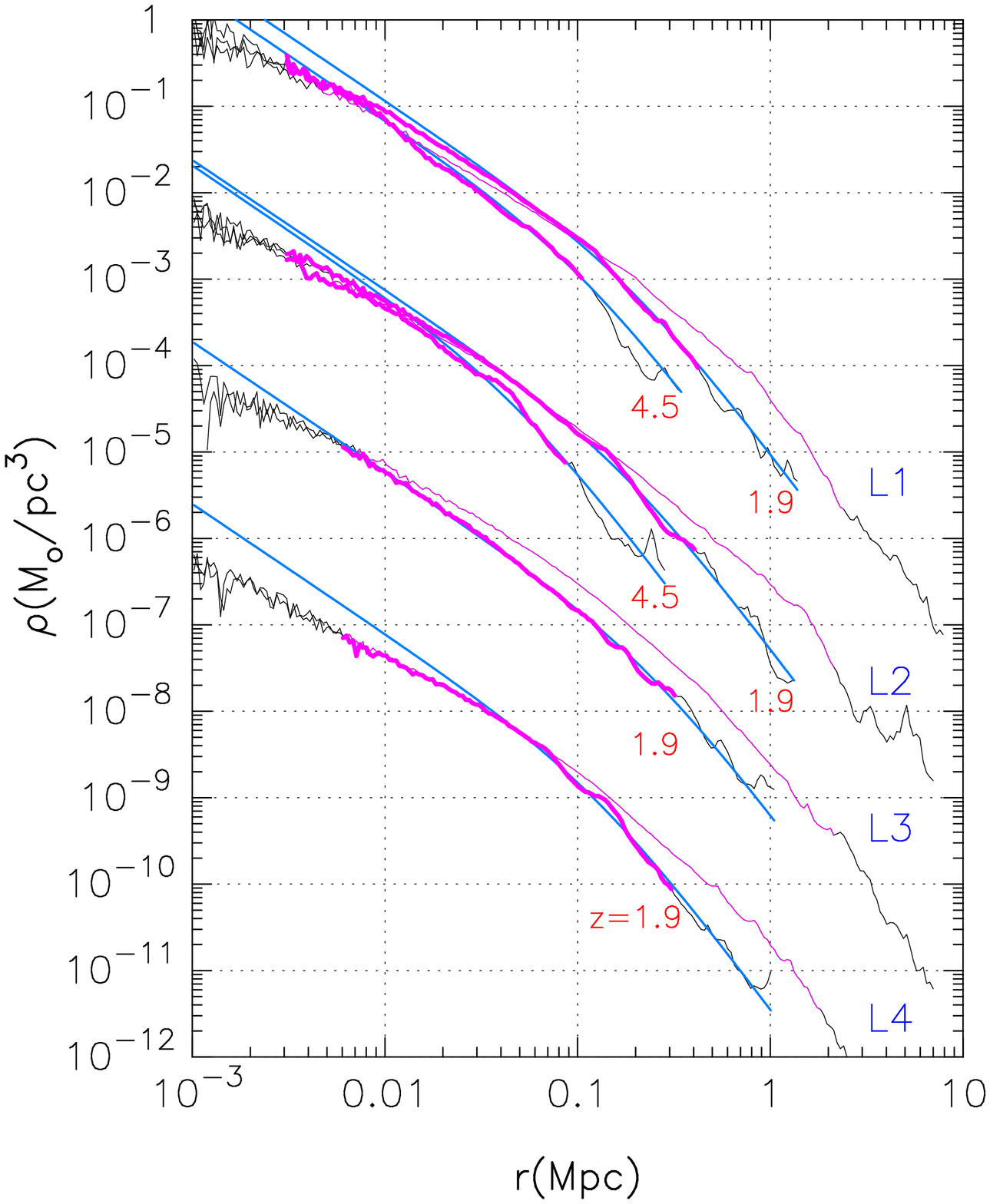}
\caption{
Same as Figure \ref{figevs}, but for the LCDM model. 
\label{figevl}}
}
\end{center}
\end{figure}

\begin{figure}
\begin{center}
{\leavevmode
\epsfxsize=15cm
\epsfbox{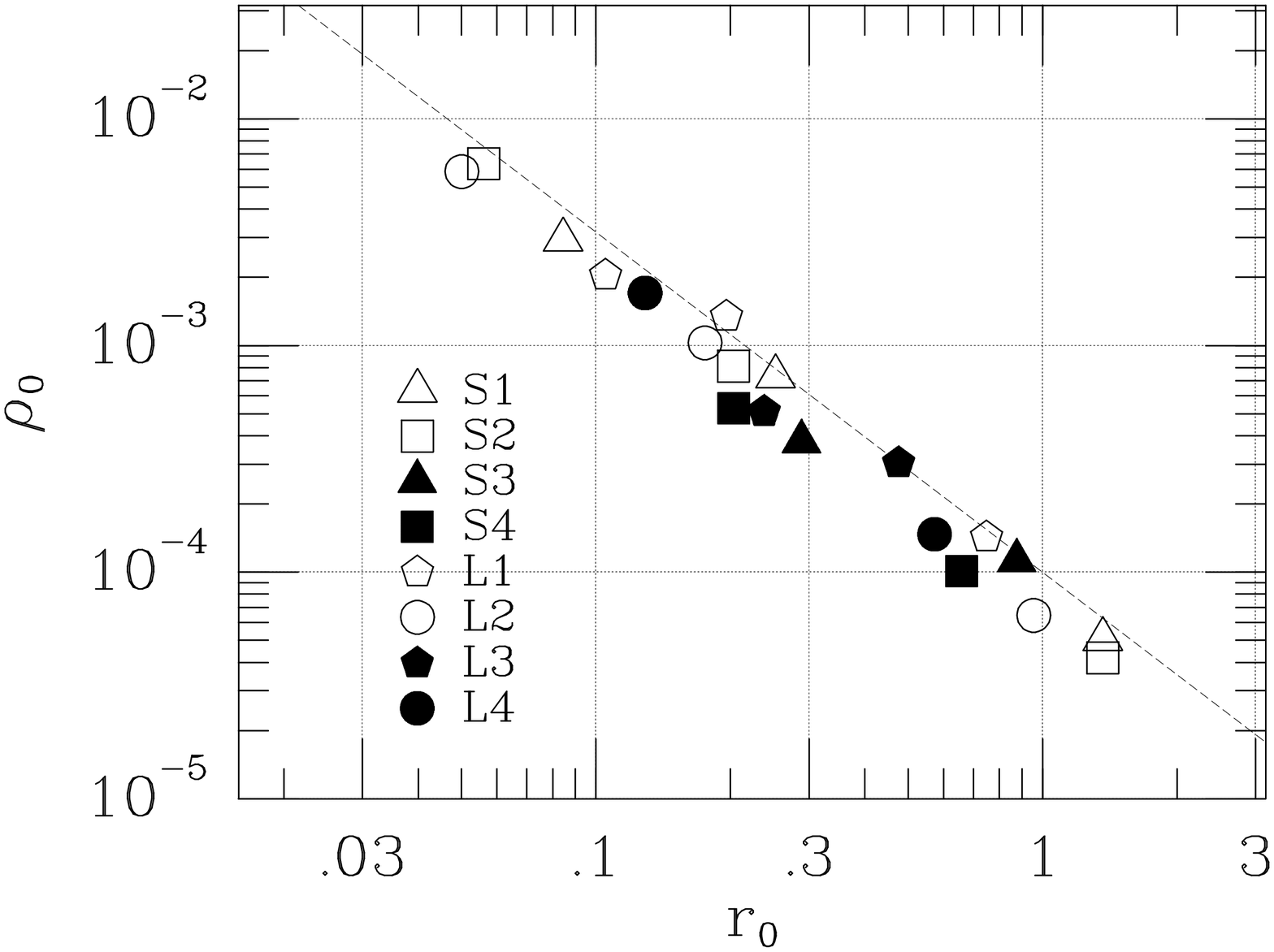}
\caption{
The scale density $\rho_0$ as a function of the scale radius $r_0$ 
(Table \ref{tab3}) at the redshift plotted in Figures \ref{figevs} and 
\ref{figevl}. The dashed line indicates $\rho_{0} \propto r_0^{-1.5}$. 
\label{figr0}}
}
\end{center}
\end{figure}

\begin{figure}
\begin{center}
{\leavevmode
\epsfxsize=15cm
\epsfbox{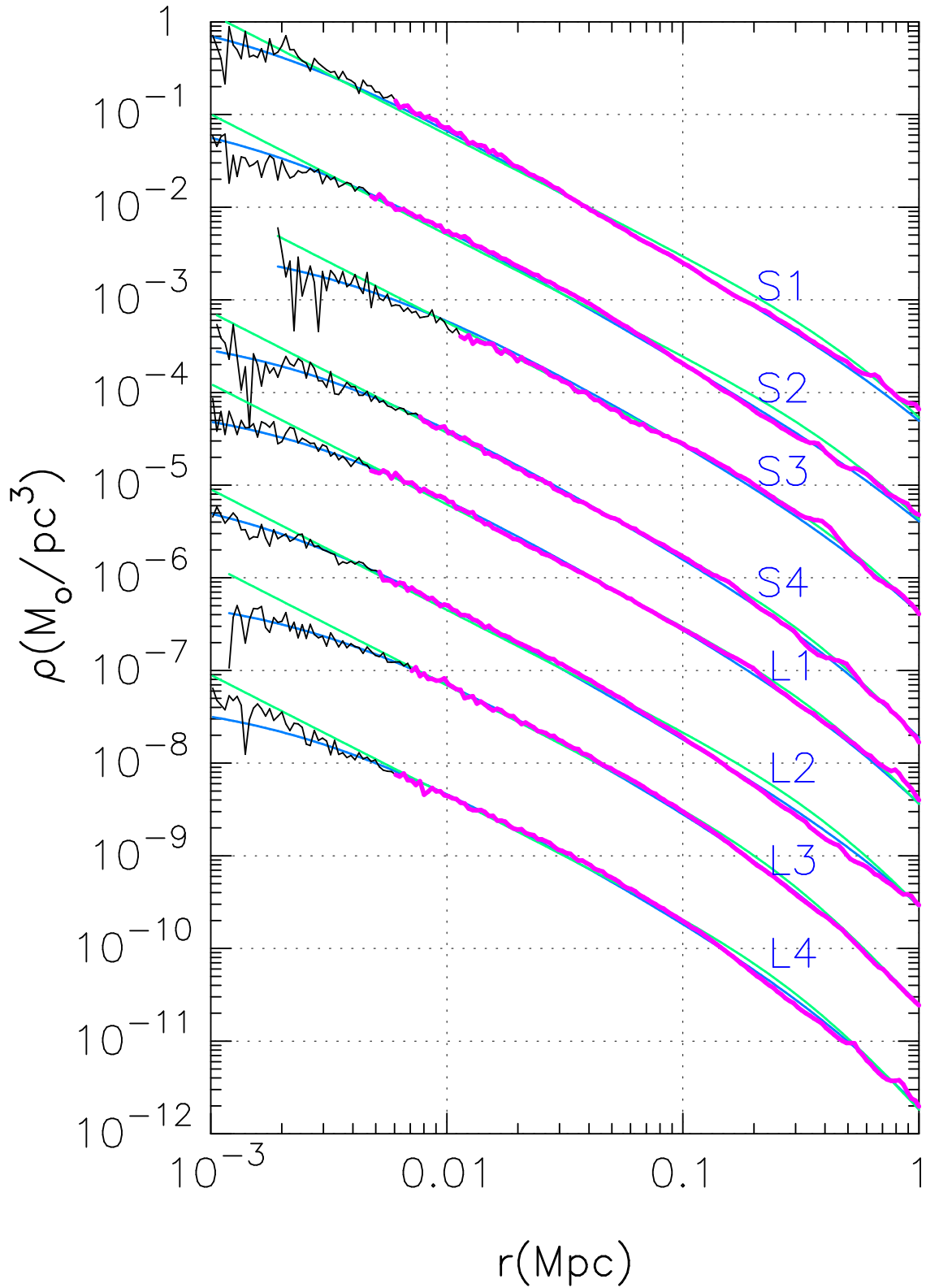}
\caption{
Density profiles for all runs at the inner region.  
The solid curves indicate the density profiles given by equations
(\ref{eqprof1}) (colored green in online edition) and (\ref{eqprof2}) 
(colored blue). 
\label{figfits}}
}
\end{center}
\end{figure}

\begin{figure}
\begin{center}
{\leavevmode
\epsfxsize=15cm
\epsfbox{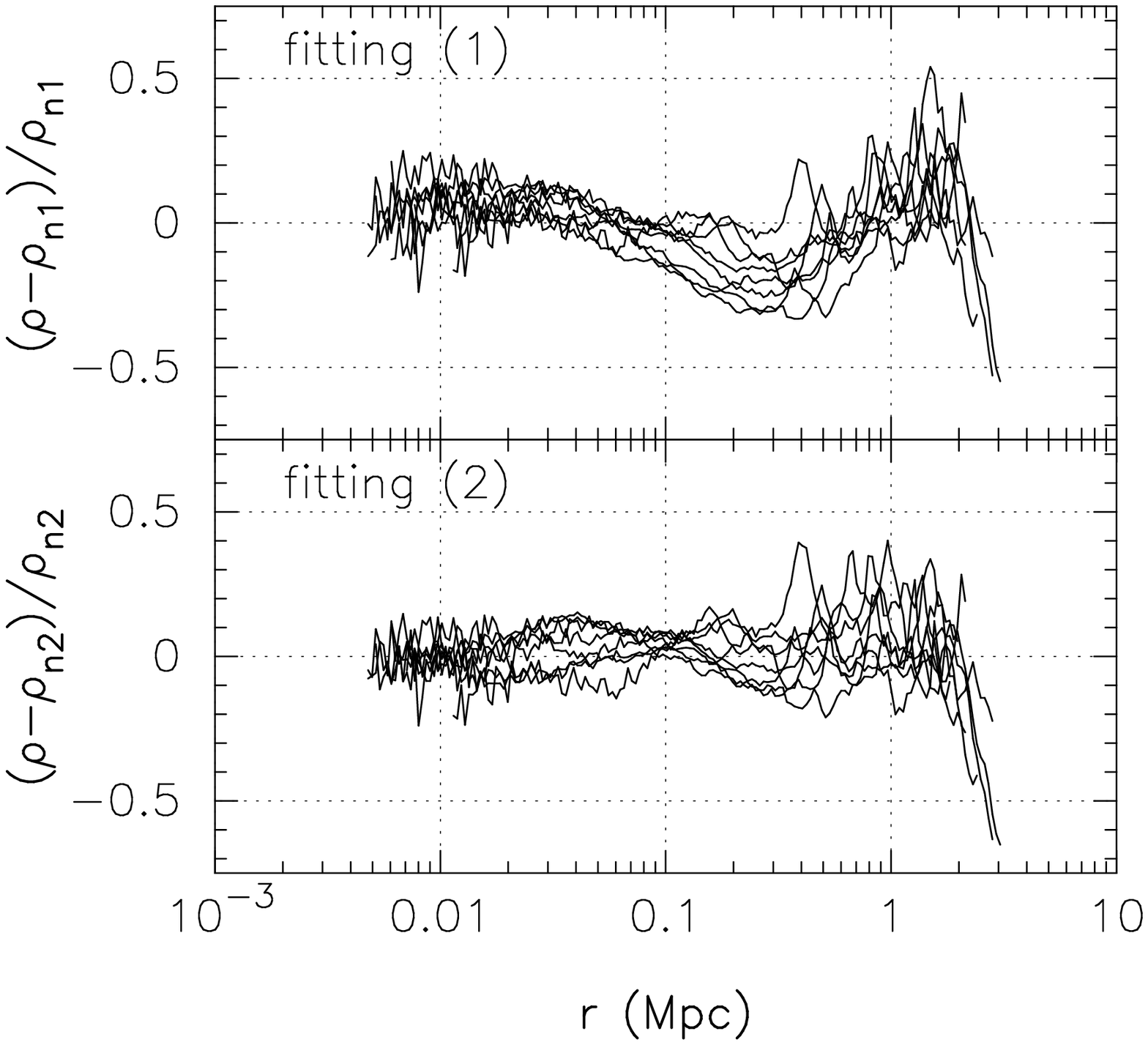}
\caption{
Residuals $(\rho-\rho_{\rm n1})/\rho_{\rm n1}$ and  
$(\rho-\rho_{\rm n2})/\rho_{\rm n2}$ 
as a function of radius. 
\label{figresnew}}
}
\end{center}
\end{figure}


\begin{references}

\reference{}
Barnes, J. E. 1990, J. Comp. Phys., 87, 161

\reference{}
Barnes, J. E., \& Hut, P. 1986, Nature, 824, 446

\reference{}
Bertschinger, E., 2001, ApJS, 137, 1

\reference{}
Eke, V. R., Cole, S., \& Frenk C. S. 1996, MNRAS, 282, 263

\reference{}
Fukushige, T., \& Makino, J. 1997, ApJ, 477, L9

\reference{}
Fukushige, T., \& Makino, J. 2001, ApJ, 557, 533 (Paper I)

\reference{}
Fukushige, T., \& Makino, J. 2003, ApJ, 588, 674 (Paper II)

\reference{}
Ghigna, S., Moore, B., Governato, F., Lake, G., Quinn, T., \&
Stadel, J. 2000, ApJ, 544, 616

\reference{}
Jing, Y. P., \& Suto, Y. 2000, ApJ, 529, L69

\reference{}
Jing, Y. P., \& Suto, Y. 2002, ApJ, 574, 538

\reference{}
Kawai, A., Fukushige, T., Makino, J., \& Taiji, M. 2000, PASJ, 52, 659

% parallel treecode on GRAPE
\reference{}
Kawai, A., Makino, J., proc. of IAU Symposium 208, 2003, in press

\reference{}
Kitayama, T., \& Suto, Y. 1997, ApJ, 490, 557

\reference{}
Klypin, A., Kravtson, A. V,, Bullock, J. S., \& Primack, J. R. 2001, 
ApJ, 554, 903

\reference{}
Makino, J. 1991, PASJ, 43, 621

\reference{}
Moore, B., Governato, F., Quinn T., Statal, J., \& Lake, G. 1998, ApJ, 499, L5

\reference{}
Moore, B., Quinn T., Governato, F., Statal, J., \& Lake, G. 1999, MNRAS, 310, 1147

\reference{}
Navarro, J. F., Frenk, C. S., \& White, S. D. M., 1996, ApJ, 462, 563 

\reference{}
Navarro, J. F., Frenk, C. S., \& White, S. D. M., 1997, ApJ, 490, 493

\reference{}
Power, C., Navarro, J. F., Jenkins, A., Frenk, C. S., White, S. D. M., Springel, V
., Stadal, J., \& Quinn, T., 
2003, MNRAS, 338, 14

% MDGRAPE-2
\reference{}
Susukita, R., Ebisuzaki, T., Elmegreen, B. G., Furusawa, H., Kato, K.,
Kawai, A., Kobayashi, Y., Koishi, T., McNiven, G. D., Narumi, T.,
Yasuoka, K.\ 2003, submitted to J. Comp. Phys.

% parallel treecode
\reference{}
Warren, M.\ S., Salmon, J.\ K., proc. of
Supercompuing '93, 1993, (IEEE Comp. Soc.), 12

\end{references}
\end{document}